\documentclass[aps,pra,preprint,superscriptaddress,longbibliography]{revtex4-2}

\usepackage{amsmath}
\usepackage{amsthm}
\usepackage{amsfonts}
\usepackage{amssymb}
\usepackage{xcolor,graphicx}
\usepackage{tikz}
\usepackage{color}
\usepackage[pdftex]{hyperref} 
\usepackage{longtable}
\usepackage{array}
\usepackage{dsfont}

\begin{document}

\title{Decoding physics identity: A Spanish-language adaptation on an instrument and its correlation with STEM achievement}

\author{O.~I. Gonz\'alez-Pe\~na }
\email[e-mail: ]{ogonzalez.pena@gmail.com}
\affiliation{Vicerrector\'ia De Ciencias de la Salud, Universidad de Monterrey, San Pedro Garza Garc\'ia, Nuevo Le\'on 66238, Mexico.}	

\author{G. Mor\'an-Soto}
\email[e-mail: ]{gmoran@itdurango.edu.mx}
\affiliation{Department of Basic Sciences, Instituto Tecnol\'ogico de Durango, Blvd. Felipe Pescador 1830, Nueva Vizcaya, Durango 34080, Dgo, Mexico.}

\author{B.~M. Rodr\'iguez-Lara}
\email[e-mail: ]{bmlara@upp.edu.mx, blas.rodriguez@gmail.com}
\affiliation{Universidad Polit\'ecnica de Pachuca, Carr. Pachuca-Cd. Sahag\'un Km. 20, Ex Hda. Santa B\'arbara, Zempoala, 43830 Hidalgo, Mexico.}	
	
\date{\today}
	
\begin{abstract}
The representation of Spanish-speaking students in STEM identity literature, particularly in physics identity, is conspicuously minimal. This study addresses this gap with a two-pronged approach. First, a physics identity instrument was adapted for Spanish-speaking STEM students to promote inclusivity in educational research. Data from 334 Mexican STEM students was collected and analysed for validation through factor analyses and Cronbach's alpha, confirming the structural integrity and reliability of the adapted instrument. Second, the instrument was employed to examine the correlation between physics identity and academic performance in a separate dataset from 200 engineering students, using multivariate linear and logistic regressions. 

The findings underscore gender's impact on physics course grades. Notably, variations in physics identity among engineering students indicated a potential link with their chosen field. Despite the importance of physics in engineering, physics identity was lower than expected, and performance in physics courses did not consistently correlate with a strong physics identity. However, the ‘interest in physics topics’ subconstruct did correlate with physics course grades.

This research fills a critical void by providing an adapted instrument for assessing physics identity in Spanish-speaking students and underscores the need for pedagogical shifts to enhance physics identity and STEM outcomes for Latino students.
\end{abstract}
	
\maketitle
\newpage

\section{Introduction}
Science, Technology, Engineering, and Mathematics (STEM) education fosters the next generation of critical thinkers and innovators who will face the world’s most urgent needs \cite{UNESCO2015}. 
As a result, there is a growing interest in studying how Sustainable Development Goals could be incorporated into STEM curricula \cite{Gonzalez2021,Rico2021}.

Developing a strong STEM identity in students is a key factor in promoting interest in STEM careers \cite{Cass2011,Godwin2016}.
A strong STEM identity creates positive expectations that may lead to engagement and success in education, career goals, and professional trajectories \cite{Tonso2006}. 
Hence, understanding and fostering the development of STEM identity becomes an important goal for educators at all levels, especially for those working with young students potentially affine to STEM majors \cite{Tonso2006}.

In education research, Basu et. al. \cite{Basu2008} defined identity as how students see themselves as powerful thinkers and doers of a specific subject; while Brickhouse et. al., \cite{Brickhouse2000} stated that students entering STEM careers often need to see themselves as the “kind of people who would want to understand the world scientifically.” 
Based on these definitions, understanding how prior experiences influence students’ STEM identity development, and how this STEM identity affects their performance once enrolled in STEM majors is of fundamental importance.

Students aspiring to be STEM professionals usually show different professional and vocational identities than their peers \cite{Capobianco2012}. 
The motivation behind the decision to pursue a STEM major may help us illustrate what kind of STEM identity these students hold prior to this choice \cite{Matusovich2011}. 
Highlighting the importance of early STEM identity development could help educators and researchers understand how and why students are drawn to STEM majors, as well as the reasons why others avoid them due to conflict within their self-perception \cite{Wang2013}.

\subsection*{Identity Theoretical Framework}
The identity framework recognizes three key subconstructs: interest, performance/competence, and recognition \cite{Godwin2016,Hazari2010}.
These three subconstructs (see Figure 1) help to understand how students perceive themselves in different fields, and they have been assessed and analysed in prior studies using qualitative \cite{Varelas2012} and quantitative \cite{Potvin2013} approaches. 
Interest positively influences students’ choice of major \cite{Tai2006}, and may increase their participation in certain activities \cite{Haussler2002}. 
This, in turn, may help students develop a positive perception of their own abilities, leading to better performance in their fields of interest. 
Performance/competence is linked to self-perception; for example, earlier performance influences future performance \cite{Marsh2002}. 
Consequently, students’ performance/competence in certain activities may influence the way they see themselves and, ultimately, their choice and persistence in majors where they can perform well \cite{Fouad1996}. 
Recognition is about how others perceive our abilities and how these perceptions affect our self-perception. 
Recognition in any field can positively influence students’ career choices \cite{Haussler2002}, and parental messages regarding their child’s abilities to perform certain activities can significantly impact on the child’s self-perception and expectations \cite{Turner2004}.

\begin{figure}
    \centering
    \includegraphics[width=0.5\linewidth]{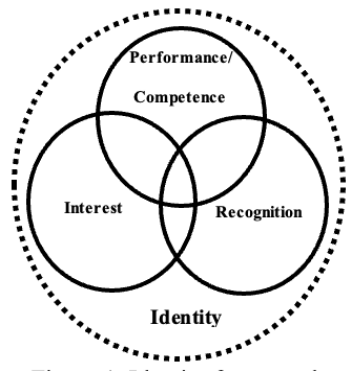}
    \caption{Identity framework.}
    \label{fig:Fig1}
\end{figure}

Following the identity framework, promoting a stronger science identity must become a priority for STEM communities and educators aiming to increase students’ interest in pursuing STEM majors around the globe. 
Moreover, promoting science-related activities at the early stages of students’ education is important because students may shift their identity within a single semester if presented with the right experiences and activities \cite{Robinson2019}. 
Increased opportunities for involvement in these kinds of activities could be of paramount importance for recruiting and training enough STEM professionals to meet the technological needs of society. 

Daily activities and home environments play a significant role in developing students’ interest in science \cite{Chen2013}; for instance, students who read science fiction and talk about science with friends and family often develop a strong science identity \cite{Dou2019} . 
Prior studies also highlight a positive correlation between science identity and students’ persistence and engagement in science-related courses such as physics and mathematics \cite{Basu2008,Brotman2008}. 
This is especially relevant for underrepresented groups in science-related careers \cite{Kahn2017}; for example, female students often struggle to develop a science identity due to gender stereotypes \cite{Chinn2002}, and female scientists with “disrupted science identities” face more career path difficulties than their male peers \cite{Carlone2007}. 
In this regard, providing educators with better tools to assess their students’ science identity could be a good starting point for designing strategies and programs to help students develop a positive picture of themselves in science \cite{Cleaves2012}; and in consequence, attract more students to science-related majors and provide them with an environment where they could persist and fulfill their goal of becoming science professionals.

\subsection*{Physics identity}
Rather than generalizing the identity framework, this research focuses on disciplinary identity, specifically the context and experiences that affect and develop physics identity. 
Unfolding physics identity into its subconstructs, research suggests that interest in physics \cite{Adams2006}, performance/competence in physics-related activities inside \cite{Marsh2002} or outside \cite{Lock2019} the classroom, as well as the recognition as a good mathematics student by family members \cite{Bleeker2004} are correlated to students’ choice of physics-related majors and their persistence in this major. 
As physics identity encompasses different dimensions of motivation \cite{Hazari2010}, it provides a lens to understand how students see themselves in the field of physics, and how they navigate their everyday experiences performing physics-related activities \cite{Enyedy2005}.

Other studies have been done on the subject of physics identity in relation to other variables, For instance, Cwik and Singh \cite{Cwik2022}  examined the impact of perceived recognition by instructors and TAs on female students' grades in physics, finding that lower perceived recognition correlates with lower grades for women. 
Patrick et al. (2018) investigated the role of engineering identity in predicting persistence, highlighting the significant influence of engineering interest, recognition, and performance/competence. 
Similarly, Prybutok et al. \cite{Prybutok2016} explored differences in engineering identity between lower and upper-division students, revealing that physics recognition is higher among upper-division students, while lower-division students exhibit higher math interest. 
While Whitcomb et al. \cite{Whitcomb2023} focused on the progression of self-efficacy, interest, identity, sense of belonging, perceived recognition, and peer interaction among physics majors, noting that perceived recognition is a key predictor of physics identity. Lastly, Verdín and Godwin \cite{Verdin2017} assessed the career intentions of first-generation engineering students, finding that those with a strong physics identity are more inclined toward non-STEM careers. 
Across these studies, the importance of identity constructs such as self-efficacy, interest, and recognition in predicting academic persistence and career choices in STEM fields is consistently underscored. 
Likewise, literature suggests that good teaching, research-based curricula, and connection to real-world applications during pre-college education improve the probability of students choosing a physics-related major \cite{StilesClarke2018}. 
However, learning environments aiming to promote the students’ interest, performance, and recognition in physics-related courses are critical to developing their physics identity \cite{Hazari2020}.

\subsection*{Women’s physics identity and the gender gap in this field}
In the field of physics, women account for approximately a quarter of awarded bachelor’s degrees \cite{Kalender2019a}. 
This underrepresentation may be caused by stereotypes that negatively influence female students’ major selection and amplify attrition rates from physics-related majors \cite{Doucette2020,Herrera2020}. 
The gender gap in physics-related majors and the professional workforce is not generated by any academic difference between males and females. 
Instead, the literature suggests that one of the main factors that negatively influence female students’ choice of these careers is the anxiety caused by the expectation of judgment based on group stereotypes \cite{Beasley2012}. 
According to current literature, female students report lower levels of physics identity than their male peers \cite{Bottomley2024,Seyranian2018} . 
A lack of recognition may hinder female students’ physics identity and their interest in physics careers, as they are less likely to feel recognized as a “physics person” \cite{Kalender2019b}.

Literature suggests that adapting physics courses and curriculum aiming to engage female students and create more interest in physics-related activities within this group could have a positive effect on these students’ physics self-concept \cite{Dawson2000}; and this could be especially important to address the poor interest reported by female students in the physical sciences at early ages \cite{Jones2000,Weinburgh1995} . 
Strong physics or science identification is usually correlated with higher grades in physics-related courses, and female students who report strong physics identity are more likely to flourish and positively evolve their performance during such courses \cite{Seyranian2018}.

Females who demonstrate talent in science reported that friends’ support was a significant predictor of their interest in pursuing a physics career \cite{Jacobs1998}. 
It is important to highlight these findings due to the significant impact that parents’ and professors’ support and recognition could have on the selection and persistence of a physics career \cite{Speering1996,Woolnough1997}. Unfortunately, parents and peers often hold gender stereotypes favouring males when comparing the abilities of males and females in science \cite{Bleeker2004,Kessels2005}.

\subsection*{The lack of language and cultural instrument adaptations to assess physics identity in Latino contexts.}
The collection of data is a fundamental step in analysing experiences designed to promote the development of students’ physics identity. 
Most instruments designed to assess identity were developed in the English language and tested for reliability and validity using American students in specific contexts; for example, the physics identity items developed by Hazari et al. \cite{Hazari2010}. 
Using a direct translation of these instruments may present students from different countries with some items that are not adapted to their context and culture \cite{Walther2014}, and their responses may create biased results and interpretation issues \cite{Temple2004}.

Some studies around the world have used direct translation of identity instruments in Mexico \cite{Oliveros2021} and other contexts \cite{Wang2020} without providing appropriate evidence of validation tests or instrument adaptations. 
Hernandez-Matias et al. \cite{HernandezMatias2019} noted that their literature review did not identify any validated Spanish language tools for assessing science identity, a trend confirmed by our own literature review.  
Direct translation jeopardizes the trustworthiness of their findings \cite{Creswell2009}. 
Therefore, it is essential to adapt the translation of instruments to the context and culture of each population and test the validity and reliability of the collected data to ensure the cultural equivalence and trustworthiness of the entire research project \cite{Chavez2005, Creswell2009}.

Both translation and cultural adaptation are equally important parts of the instrument adaptation process. 
Failing to thoroughly complete these two parts may result in a deficient instrument, jeopardizing the reliability of the collected data.

\section{Purpose}
The goal of this research is to expand current literature on physics identity within Latino contexts and Spanish-speaking populations, aiming to promote greater interest in this topic from researchers worldwide. To achieve this goal, the research focuses on the following objectives:
\begin{enumerate}
    \item Adapt an existing instrument to assess the physics identity of STEM Mexican students in the Spanish language and test the validity of the data collected with this instrument. 
    \item Analyse the correlation between STEM students’ physics identity and their grades in physics-related courses. 
\end{enumerate}

By adapting this instrument to collect more accurate assessments of physics identity among students in the 
Mexican context, we aim to expand the tools available to Spanish-speaking STEM educators. 
This adaptation is especially important for future research projects analysing physics identity in Latino populations, as it will facilitate data collection and validation while considering the cultural equivalence of students from different countries, cultures, and contexts \cite{Walther2014}.

\section{Methods}
The first part of this study relies on a mixed-methods approach to adapt an existing instrument aiming to collect and analyse reliable data about Mexican STEM students’ physics identity. Data collected from STEM students of two northern Mexican institutions were tested for content validity and reliability using factorial analyses and Cronbach’s alpha for the instrument adaptation. For the second part of this study, an additional dataset was collected to conduct a logistic binomial regression and a multivariate linear regression aiming to analyse how students’ physic identity could influence their performance in physics courses. All statistical analyses were conducted using the statistical software R \cite{Team2012}.

\subsection*{Survey Instrument}
A section of the instrument used to collect data for the Persistence Research in Science and Engineering (PRiSE) Project was translated from English to Spanish language and some adaptations were made to better reflect the Mexican STEM students’ context. 
This instrument, developed by the Science Education Department of the Harvard-Smithsonian Center for Astrophysics, aims to identify upper high school factors that could influence the persistence of U. S. college students in STEM majors in 2007 (the survey can be viewed online at \url{https://www.cfa.harvard.edu/sed/projects/PRiSE_survey_proof.pdf}). 
A total of 28 Likert-type items were selected from the PRiSE questionnaire as these items where originally developed to assess the construct of physics identity. 
These 28 items rate students’ self-perceptions about their physics identity on a scale ranging from 0 (“No, not at all”) to 5 (“Yes, very much”). 
These items were designed aiming to assess four dimensions or subconstructs of physics identity: Recognition (2 items), Performance (5 items), Competence (2 items), and Interest (19 items; with 6 items for the physics interest subconstruct, and 13 items for the science interest subconstruct).
The average of these four subconstructs is calculated to determine a person’s physics identity. 
The PRiSE questionnaire was selected for this adaptation due to the validity evidence reported by previous research conducted with the identity items of this instrument \cite{Godwin2016,Hazari2010}.

\subsection*{Instrument Adaptations}
There is evidence suggesting that the identity subconstructs of competence and performance in the original instrument could be considered as one subconstruct: “Competence/Performance” \cite{Godwin2016,Hazari2010}. 
This combined subconstruct could help educators and researchers to better understand the variables of previous performance in physics courses, as the items of these subconstructs are very similar and usually load together in exploratory factor analyses testing students’ identity in different contexts \cite{Godwin2016,Hazari2010}. 
Following these previous research suggestions, the subconstruct competence/performance was hypothesized as one singular identity subconstruct during the adaptation and validation process.

An additional adaptation was made to the Likert-type scale of the PRiSE items, selecting a seven-point anchored scale (from 0 – “strongly disagree” to 6 – “strongly agree”) to replace the Likert-type scale. 
This scale was modified aiming to facilitate the interpretation of the distance between each numeric response. 
For example, the distance between an item response of 0 and 1 on an anchored scale between 0 and 6 is the same as the distance between a response of 1 and 2, and this distance is more objective than the distance between a response of “somewhat agree” and “agree”.

\subsection*{Translation and Context Adaptation}
The 28 items assessing physics identity were translated from English to Spanish by five different STEM professors. 
These professors were fluent in Spanish and English and had experienced both the U.S. and Mexican educational contexts and cultures. 
They had teaching experience in mathematics, physics, and chemistry courses, as well as experience conducting education research in both countries. 
The five professors translated the 28 items individually, and they compared and discussed their translation and personal interpretation of the items aiming to determine the most appropriate way to translate each item to the Spanish language. 
This process resulted in a Spanish language translation that kept the meaning and essence of the original items and could be easily understood by Mexican STEM students \cite{Chavez2005}.

Aiming to avoid possible cultural and context misunderstandings with the translated items, three focus groups were conducted with STEM students and professors. The first focus group participants were four physics professors from one of the universities where this research was conducted, the second focus group was conducted with four students from the same university, and the third focus group was conducted with three students from the other university where this research was conducted. 
All focus group participants received a copy of the translated instrument a day prior to their focus group with instructions to read the entire instrument and reflect on possible misunderstandings or lack of clarity about the items. 

During the focus groups, the participants discussed about possible issues with the clarity of some items. The feedback from these focus groups helped to establish content and face validity \cite{Creswell2009,DeVellis2016}. 
The cultural validation process is as important as the language translation. 
As a result, some adaptations were made to the wording of some items trying to ensure that Mexican STEM students would interpret the items correctly. 
For example, one item asked whether parents/relatives/friends typically see the student as a physics person, and that wording could be confusing. 
According to the focus groups’ feedback, Mexican students may feel that this item is asking two different questions at the same time because parents and family are very important to Mexican students. 
Hence, friends should be considered as a different group of people that could create a different response to this item. Following these comments, the item was separated into two different questions. 

In the end, the final Spanish language adaptation ended with 29 items with an acceptable cultural equivalence of the items \cite{Andrews2015,Chavez2005} that will avoid misinterpretations collecting data from Mexican STEM students. 

\section*{Participants and Data Collection}
Two datasets were collected in two different semesters, one during the Fall 2020 and the second one during the Spring 2021. 
The first dataset was collected using the physics identity instrument adaptation in two northern STEM universities in Mexico. 
One of the universities is the most important private institution for technological education in the country, and the other university is part of the largest engineering public education system in Mexico. 
A total of 334 participants taking physics-related courses such as introduction to physics, dynamics, and statics from both universities, 153 (46\%) participants from the private university, and 181 (54\%) from the public university completed the physics identity instrument for the Fall 2020 dataset. 
Most of the participants were male, with 248 (74\%) male students and 86 (26\%) female students. The Spring 2021 dataset was collected using the instrument adaptation that resulted from the validation tests. 
This validated instrument was distributed only in the public Mexican university; collecting information about the physics identity of 200 participants enrolled in different engineering majors, with 122 (61\%) males and 78 (39\%) females. 
Additionally, to the physics identity information, the final grades of the physics-related courses of all participants from the Spring 2021 dataset were collected at the end of the semester. 
The physics-related course grades ranged from 70 to 100 for all passing grades, and the failing grades were reported as a 60 since all failing grades are reported as a “failed or F” in students’ report cards at the end of the semester. All the participants from the two datasets were enrolled in second or third-semester STEM majors and were taking a mandatory physics-related course for their major. 

The Fall 2020 dataset was used to conduct validity tests for the adaptation of the instrument to assess Mexican STEM students’ physics identity. 
This sample (n = 334) was divided into two halves (n = 167) to conduct a different factorial analysis with each half of the data. 
There are different theories about the appropriate sample size for running factor analysis. 
For instance, some authors cite a minimum of 100 subjects to conduct any factor analysis \cite{Gorsuch1983,Kline1994}; while others, like Cattell \cite{Cattell1978}, suggest three to six subjects per variable. 
Our sample size of 167 participants for each factorial analysis is above the suggested minimums and in line with Cattell’s suggestion with a subject (n = 167) to item (29) radio of 5.7 for each analysis. 
The Kaiser-Meyer-Olkin (KMO) measure of sampling adequacy was checked for both halves of the sample \cite{Kaiser1970}, with a KMO=0.88 for the first half (exploratory factor analysis) and a KMO = 0.92 for the second half (confirmatory factor analysis). 
According to the KMO values for each split-half dataset, there is a good correlation pattern, indicating that our sample size is adequate for the quantitative analysis \cite{Hutcheson1999}.

We examined the univariate normality of the data using skew and kurtosis. Absolute values of 2.0 or higher for skewness and 7.0 or higher for kurtosis are expected for a normal distribution \cite{Curran1996}. 
Additionally, we checked the multivariate normality of the data using Mardia’s coefficient. If Mardia’s coefficient is significant with a p-value below 0.05, then the multivariate normality is violated for the data set \cite{Curran1996}.

\subsection*{Exploratory Factor Analysis}
The exploratory factor analysis (EFA) is a data-driven approach that relies on the common factor model with the goal of inductively analysing the possible relationships among all items to determine if an underlying structure exists \cite{Fabrigar1999}. 
The EFA test is if each variable in a group of measured variables is a linear function of one or more common factors. These common factors are latent (unobservable), and they cannot be directly measured. 
Therefore, the EFA is usually used during the development of new scales where the researchers do not know the underlying pattern of the new items \cite{Lee2018}.

One of the underlying assumptions of factor analysis is the normality of the data \cite{Fabrigar1999}. 
If the data normality is proved, then a maximum likelihood factor analysis is the best fit; but if this assumption is severely violated, then the model-fitting procedure can produce inaccurate results \cite{Curran1996}. 
Our dataset showed items that indicated a univariate non-normal distribution with skewness ranging from -0.17 to -1.62, and kurtosis values from 1.91 to 5.52, with a significant Mardia’s coefficient p-value $< 0.001$ indicating violation of multivariate normality. 
Therefore, we conducted an EFA with the first half of the data (n = 167) using an ordinary least squares estimator aiming to find the minimum residual solution. 
This is a maximum likelihood with robust standard errors EFA that produces similar results to the maximum likelihood estimator solution when the data shows a non-normal distribution \cite{Revelle2014}. 
A promax (nonorthogonal or oblique) rotation was selected to allow inter-correlations between the constructs. 
This oblique rotation is normally used in social science research because it is a more realistic representation of the measured phenomena \cite{Fabrigar1999}, and we were expecting a strong correlation between the constructs based on our framework and previous studies about physics identity \cite{Hazari2010}. 
A rule of thumb of 0.32 or higher was used as the minimum loading factor for an item to be considered as a part of a factor, which equates to approximately 10\% overlapping variance with the other items in that factor \cite{Tabachnick2001}.

A scree plot and a parallel analysis were used to determine the appropriate number of factors to conduct the EFA. 
The scree plot visual result suggested a four factors EFA, and this information was supported by the parallel analysis results and the number of subconstructs that the original instrument presented and hypothesized for assessing multiple dimensions of the physics identity. 
Based on these results, the EFA was run with four factors: 1) Recognition, 2) Competence/Performance, 3) Physics interest, and 4) Science interest.

\subsection*{Confirmatory Factor Analysis}
A confirmatory factor analysis (CFA) was conducted with the other half of the data (n = 167) following the EFA results. 
The CFA tests if the items hypothesized to measure a single underlying latent construct based on the EFA results are, in fact, doing that \cite{Brown2015}. 
The CFA explicitly tests a model specified by the researcher where items load onto a specific factor rather than letting the factor loadings emerge from the data analysis like the EFA \cite{Fernandez2010}. 
A maximum likelihood factor analysis was used, selecting to fix the factor variance to one and allowing the standardized estimates of each path to be estimated for this analysis. 
Additionally, the latent factors were allowed to covary, which is consistent with the oblique rotation that was previously used in the EFA \cite{Fabrigar1999,Lee2018}.

The model fit and significance of the paths between latent variables were tested for the four constructs (recognition, competence/performance, physics interest, and science interest) with the CFA using the 23 remaining items according to the EFA results. 
Several fit indices were used for evaluating the CFA model fit including the chi-square, which should be non-significant at the p > 0.05 value \cite{Byrne1994}; the Comparative Fit Index (CFI), expecting acceptable values above 0.95 \cite{Brown2015,Hu1999}; the Tucker Lewis Index (TLI), expecting acceptable values above 0.95 \cite{Brown2015,Hu1999}); and the root mean square error of approximation (RMSEA), expecting values less than 0.08 for moderate fit, 0.05 for a good fit, or 0.01 for excellent fit \cite{MacCallum1996}. 
All these different types of fit indexes were used because each gives a different facet of model fit, providing multiple sources of information that must be triangulated to determine how well the specified model fits the data.
Additionally, all the fit indexes and chi-square statistics of the CFA were adjusted using the Satorra-Bentler correction aiming to find a robust solution to non-normal data \cite{Satorra1988}. 
Aiming to optimize model fit indices and statistical significance in the CFA results, some of the items that showed the lowest EFA factor loadings were deleted one by one, and additional CFA was conducted after each item deletion \cite{BatistaFoguet2004}.

\subsection*{Internal Consistency}
The 14 remaining items after the validity tests through factor analyses were tested for internal consistency using Cronbach's alpha ($\alpha$), expecting values of 0.80 or higher for each of the four physics identity subconstructs \cite{Thorndike2010}.

\subsection*{Linear Regressions}
The Spring 2021 dataset was used to analyse which variables affected participants’' grades in their physics-related courses using both, a multivariate linear regression model \cite{Collett1991,Hayes2013}, and a logistic binomial regression model \cite{Hayes2013}. 
These regression models were conducted using the physics-related course grades as an independent variable. 
The multivariate linear regression model used the physics-related course grade ranging from 60 to 100 (this range established all failing grades as 60 and all passing grades between 70 and 100) as the dependent variable with six independent variables: X1) Recognition, X2) Competence/Performance, X3) Physics interest, X4) Science interest, X5) Physics-related course, and X6) Gender. 
For the logistic binomial regression, the Spring 2021 data was classified into two different groups using the participants' physics-related course grades as a dichotomous variable \cite{Creswell2009}, separating participants with failing grades in one group, and participants with grades equal to or greater than 70 in the other group. 
These two groups (failing and passing grades) were used as a dichotomous dependent variable for the logistic binomial regression model, and the same six independent variables used in the multivariate linear regression were considered for this model: X1) Recognition, X2) Competence/Performance, X3) Physics interest, X4) Science interest, X5) Physics-related course, and X6) Gender.

\section{Results}

\subsection*{Exploratory Factor Analysis}
The results for the EFA are shown in Table I. Following the EFA results six items were removed from three of the four subconstructs of the instrument adaptation: one from competence/performance (Q7), two from physics interest (Q15 \& Q16), and three from science interest (Q26, Q27 \& Q29). These six items were removed because they did not fulfil the minimum EFA loading factor of 0.32 for any subconstruct. The four-factor model used for the EFA accounted for 54\% of the variance for the analysed items, which is within the accepted ranges for new instruments in social science research \cite{Hair2014,Peterson2000}.

\begin{table}[h!]
\centering
\caption{Loading factors of the exploratory factor analysis.  Items indicated with a * were ultimately removed from the CFA to strengthen the model fit.}
\label{tab:Table1}
\begin{tabular}{ l c c c c }
\hline
\textbf{Item} & \textbf{Recognition} & \textbf{Competence/Performance} & \textbf{Physics Interest} & \textbf{Science Interest} \\ \hline
Q1  & 0.82 &       &       &       \\ 
Q2  & 0.88 &       &       &       \\ 
Q3  & 0.54 &       &       &       \\
Q4* &      & 0.70  &       &       \\ 
Q5  &      & 0.87  &       &       \\ 
Q6* &      & 0.56  &       &       \\ 
Q8* &      & 0.55  &       &       \\ 
Q9  &      & 0.82  &       &       \\ 
Q10 &      & 0.88  &       &       \\ 
Q11 &      &       & 0.91  &       \\ 
Q12 &      &       & 0.76  &       \\ 
Q13 &      &       & 0.77  &       \\ 
Q14* &     &       & 0.66  &       \\ 
Q17* &     &       &       & 0.51  \\ 
Q18* &     &       &       & 0.74  \\ 
Q19 &      &       &       & 0.90  \\ 
Q20 &      &       &       & 0.80  \\ 
Q21* &     &       &       & 0.34  \\ 
Q22 &      &       &       & 0.86  \\ 
Q23 &      &       &       & 0.96  \\ 
Q24 &      &       &       & 0.81  \\ 
Q25* &     &       &       & 0.63  \\ 
Q28* &     &       &       & 0.55  \\ \hline
\end{tabular}
\begin{flushleft}
\end{flushleft}
\end{table}

\subsection*{Confirmatory Factor Analysis}
A CFA was conducted with the second half of the data (n = 167) using the 23 remaining items after analysing the EFA results (see Table I). 
The CFA model fit did not achieve acceptable standards with the 23 items, and items with the lowest factor loadings in the EFA results were systematically removed aiming to achieve the acceptable model fit \cite{Brown2015,Hu1999} (see Table I). 
Following this approach, nine additional items were removed from three subconstructs of the adapted instrument: three from competence/performance (Q4, Q6 \& Q8), one from physics interest (Q14), and five from science interest (Q17, Q18, Q21, Q25 \& Q28). 
The final model tested with the CFA kept 14 items (see Table I); Figure 2 shows the Spanish language final version of the instrument adaptation. 
The CFA conducted with these 14 items showed good overall fit indices, with a Comparative Fit Index of 0.967, a Tucker Lewis Index of 0.958, and a root mean square error of approximation value of 0.056. 
The chi-square statistic was significant at a p $< 0.001$, which was an expected result due to the sample size (n = 167), but it is still considered a good model fit. 
These values for the fit indexes suggest that the model shown in Figure 3 is appropriate for data collection with similar populations, providing valid evidence for the items assessing the physics identity of Mexican STEM students.

\begin{figure}
    \centering
    \includegraphics[width=\linewidth]{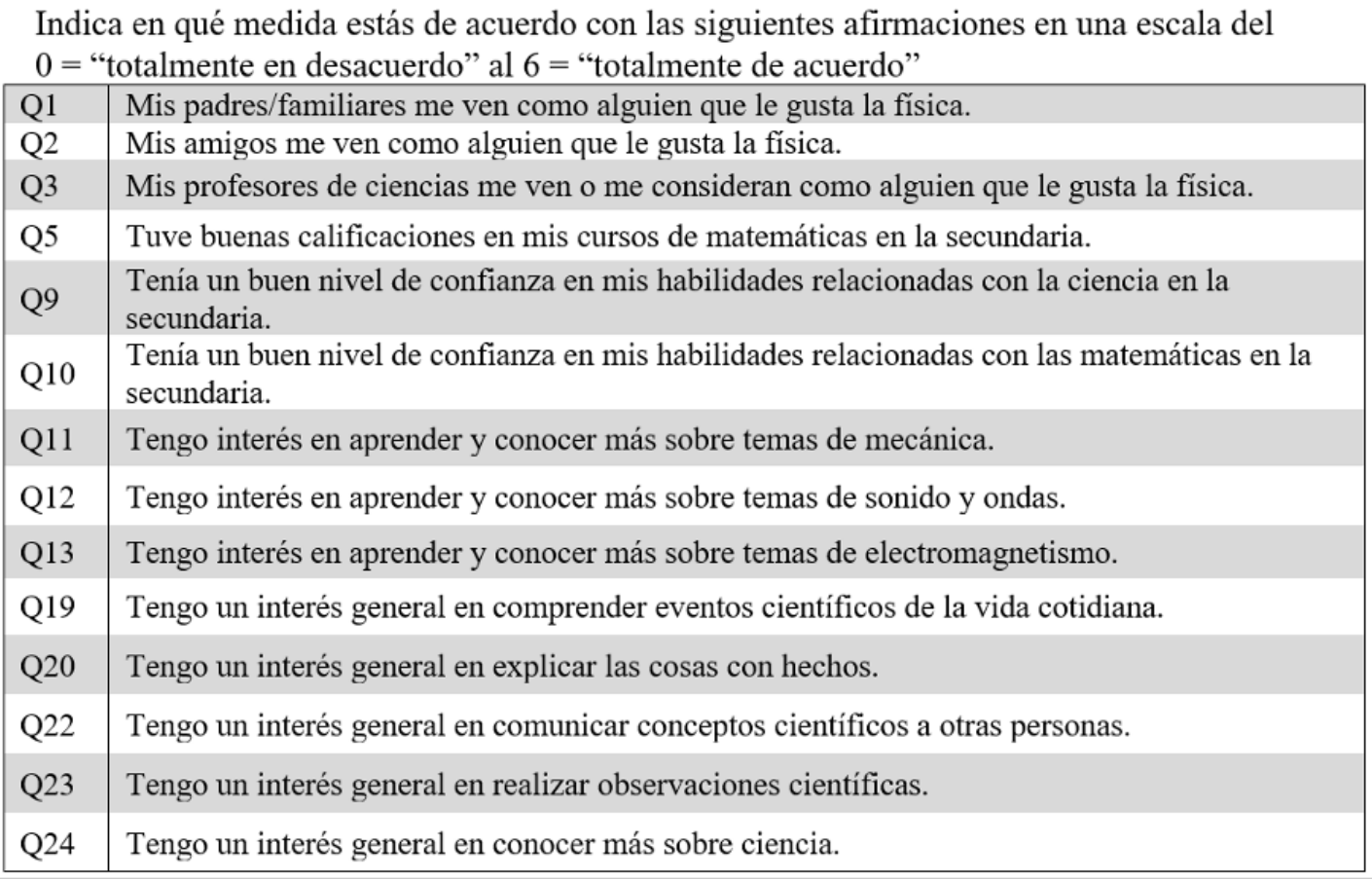}
    \caption{Final version of the instrument to assess physics identity of Mexican STEM students.}
    \label{fig:Fig2}
\end{figure}

\begin{figure}
    \centering
    \includegraphics[width=0.75\linewidth]{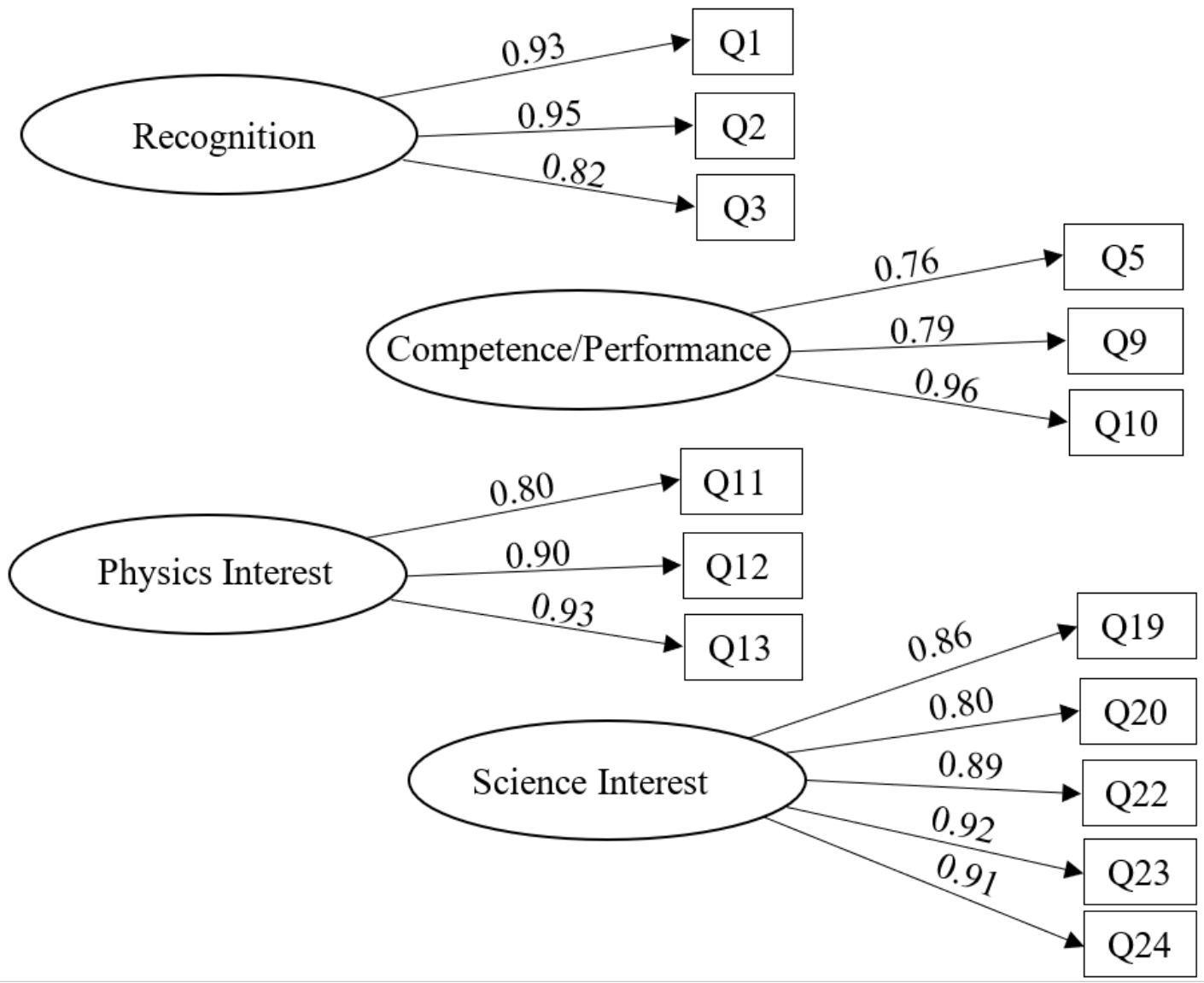}
    \caption{Model of resulting CFA. All paths shown are significant at the $p < .001$ level}
    \label{fig: Fig3}
\end{figure}

\subsection*{Internal Consistency}
The Cronbach’s alpha values for the four subconstructs showed strong internal reliability with coefficients of $\alpha = 0.88$ for recognition; $\alpha = 0.86$ for competence/performance; $\alpha = 0.89$ for physics interest; and $\alpha = 0.93$ for science interest. These alpha values provided enough evidence to determine that the remaining items measure each subconstruct as expected \cite{Hair2014}.

\subsection*{Participants' Performance in Physics-Related Courses}
The information about participants’ gender, major, physics-related course grades, and their physics identity is presented in Table II.

\begin{table}[h!]
\centering
\caption{Participants from each major and their physics-related course information.}
\label{tab:Participants}
\begin{tabular}{l c c c c c c c }
\hline
\textbf{Engineering Major} & \textbf{Participants} & \textbf{Male} & \textbf{Female} & \textbf{Passed} & \textbf{Failed} & \textbf{Physics Grade} & \textbf{Physics Identity} \\ \hline
Business management        & 73                   & 25           & 48             & 67             & 6              & 88.23                 & 3.66                      \\ 
Electric                   & 8                    & 7            & 1              & 7              & 1              & 78.88                 & 3.54                      \\ 
Electronic                 & 8                    & 7            & 1              & 8              & 0              & 87.50                 & 3.64                      \\ 
Industrial                 & 9                    & 6            & 3              & 9              & 0              & 84.00                 & 3.83                      \\ 
Mechatronic                & 23                   & 19           & 4              & 15             & 8              & 76.65                 & 4.43                      \\ 
Civil                      & 57                   & 44           & 13             & 46             & 11             & 81.40                 & 3.65                      \\ 
Mechanical                 & 15                   & 12           & 3              & 8              & 7              & 74.53                 & 4.07                      \\ 
Chemical                   & 7                    & 2            & 5              & 6              & 1              & 83.43                 & 4.27                      \\ \hline
\textbf{Total}             & \textbf{200}         & \textbf{122} & \textbf{78}    & \textbf{166}   & \textbf{34}    & \textbf{83.17}         & \textbf{3.80}             \\ \hline
\end{tabular}
\end{table}

\subsection*{Multivariate Linear Regression}
Table III shows the results of the multivariate linear regression pinpointing the independent variables that were significant predictors of participants' physics-related course grades.

\begin{table}[h!]
\centering
\setlength{\tabcolsep}{30pt}
\caption{Multivariate linear regression model results.}
\label{tab:regression}
\begin{tabular}{l c c c}
\hline
\textbf{Independent Variables} & \textbf{B}   & \textbf{SE}   & \textbf{\textit{p}-Value} \\ \hline
Recognition                    & 0.116        & 0.729         & NS                        \\ 
Competence/Performance         & 0.853        & 0.671         & NS                        \\ 
Physics interest               & -2.349       & 0.960         & 0.015*                    \\ 
Science interest               & 1.841        & 1.057         & NS                        \\ 
Gender                         & -3.428       & 1.830         & NS                        \\ 
Physics-related course         & -1.207       & 0.356         & $<0.001$***                 \\ \hline
\end{tabular}
\begin{flushleft}
B = estimated coefficient; SE = standard error; NS = not significant.
\end{flushleft}
\end{table}

After further analysis of the Table III, an additional multivariate linear regression model was conducted eliminating the physics-related course independent variable from the model aiming to better understand the effect that the other five independent variables could have in the participants’ physics-related course grades. The results of this new multivariate linear regression model are presented in Table IV.

\begin{table}[h!]
\centering
\setlength{\tabcolsep}{30pt}
\caption{Results of the multivariate linear regression model without the physics-related course variable.}
\label{tab:regression_no_course}
\begin{tabular}{ l c c c }
\hline
\textbf{Independent Variables} & \textbf{B}   & \textbf{SE}   & \textbf{\textit{p}-Value} \\ \hline
Recognition                    & -0.013       & 0.747         & NS                        \\ 
Competence/Performance         & 0.906        & 0.688         & NS                        \\ 
Physics interest               & -2.376       & 0.986         & 0.016*                    \\ 
Science interest               & 1.692        & 1.085         & NS                        \\ 
Gender                         & -5.322       & 1.789         & 0.003**                   \\ \hline
\end{tabular}
\begin{flushleft}
B = estimated coefficient; SE = standard error; NS = not significant.
\end{flushleft}
\end{table}

\subsection*{Logistic Binomial Regression}
The results shown in Table V provide information about the factors that could influence students' performance in their physics-related courses, using a passing or failing grade as a dichotomous independent variable.

\begin{table}[h!]
\centering
\setlength{\tabcolsep}{30pt}
\caption{Logistic binomial regression model results.}
\label{tab:logistic_regression}
\begin{tabular}{ l c c c }
\hline
\textbf{Independent Variables} & \textbf{B}   & \textbf{SE}   & \textbf{\textit{p}-Value} \\ \hline
Recognition                    & -0.101       & 0.174         & NS                        \\ 
Competence/Performance         & 0.167        & 0.155         & NS                        \\ 
Physics interest               & -0.350       & 0.262         & NS                        \\ 
Science interest               & 0.289        & 0.277         & NS                        \\ 
Gender                         & -0.869       & 0.502         & NS                        \\ 
Physics-related course         & -0.231       & 0.095         & 0.015*                   \\ \hline
\end{tabular}
\begin{flushleft}
B = estimated coefficient; SE = standard error; NS = not significant.
\end{flushleft}
\end{table}

Similarly to the multivariate linear regression model, an additional logistic binomial regression model was conducted removing the physics-related course independent variable. The results of this new logistic binomial regression model are presented in Table VI.

\begin{table}[h!]
\centering
\setlength{\tabcolsep}{30pt}
\caption{Results of the logistic binomial regression model without the physics-related course variable.}
\label{tab:logistic_regression_no_course}
\begin{tabular}{ l c c c }
\hline
\textbf{Independent Variables} & \textbf{B}   & \textbf{SE}   & \textbf{\textit{p}-Value} \\ \hline
Recognition                    & -0.130       & 0.168         & NS                        \\ 
Competence/Performance         & 0.166        & 0.152         & NS                        \\ 
Physics interest               & -0.296       & 0.251         & NS                        \\ 
Science interest               & 0.208        & 0.260         & NS                        \\ 
Gender                         & -1.148       & 0.487         & 0.018*                   \\ \hline
\end{tabular}
\begin{flushleft}
B = estimated coefficient; SE = standard error; NS = not significant.
\end{flushleft}
\end{table}

\section{Discussion}

\subsection*{Instrument Adaptation}
The results of the first objective of this research are presented as the adaptation and validation process of the instrument to assess physics identity in Latino contexts. 
These results suggest that the adaptation of this instrument could facilitate the collection of reliable data to conduct future research about the physics identity of Spanish-speaking STEM students (see the instrument in Figure 2). 
The adaptation process presented in this research highlights the importance of conducting validation tests on any dataset collected with new or adapted instruments to improve the quality and trustworthiness of a research project. 
This adaptation process considered the possible language barriers and cultural differences between different populations \cite{Walther2013}, which is of paramount importance when conducting research based on data collected using adaptations of existing instruments in different contexts. 
Additionally, to the language and cultural adaptations, several psychometric tests were performed to determine the final version of the instrument to assess physics identity. 
This whole process could facilitate the design of physics identity research in other Spanish-speaking Latino contexts, and populations that may feel a better connection with the language and context presented in the adapted items \cite{Walther2014}, and further validation tests could be performed on datasets collected from different Latino populations using this instrument adaptation as a baseline. 
In the end, having new instruments that better reflect the language and cultural differences of different populations could help researchers to design more thorough studies about physics identity, which ultimately could help to develop a deeper interest in this topic from researchers around the world that may feel more included in the global STEM research agenda.

After the validation tests 15 items were removed from the instrument adaptation. 
The only subconstruct that did not lose any item was recognition, suggesting that recognition from family, friends, and science professors as a physics person is important for students’ physics identity development in the Latino populations. 
The competence/performance subconstruct lost four items; all these items were related to students’ grades and performance in specific science and math courses like the first physics course in upper high school. 
It is possible that Latino students perceived these courses as different in content and difficulty (science, math, and physics), and they answered those items in a very different way. 
It is important to highlight that the only item related to students’ prior performance and grades that the competence/performance subconstruct kept after the validation process was the middle school math course grades. 
This item loaded into this subconstruct with two items related to students’ confidence level performing science and math activities, suggesting that math self-efficacy and math performance could be significant factors in Latino students’ physics identity development. This finding is in line with prior literature suggesting that math preparation is positively correlated to students’ physics identity development and interest in physics-related courses \cite{Bleeker2004, Chen2021,Lock2019,Meltzer2002} . 
Two of the three deleted items from the physics interest subconstruct were related to courses that are not part of the science courses offered by Latino upper high schools (relativity and physics history); this may be the reason why these items did not load into this subconstruct with the items asking about interest in topics that these students usually discuss in their physics courses in upper high school like mechanics, waves, and electromagnetism. 
Regarding the science interest subconstruct, the five items that were kept after the validation process were related to activities that are usually performed and practiced in a school environment like laboratories, classes, and discussions with peers and professors, while most of the six deleted items were related to students’ interest in activities that are usually performed as extracurricular activities or hobbies in Latino contexts such as science clubs, reading science fiction, or science competitions. 
The deleted items from the science interest subconstruct could have been perceived by Latino students as optional extracurricular activities that only students with a deep interest and involvement with science topics want to pursue. 
This perception about the deleted items could have created a diverse set of Latino students’ responses that led to the elimination of these science interest subconstruct items during the validation process.

\subsection*{Physics Identity Levels of Latino STEM Students}
The results of the second objective of this research show that most of the participants of this study reported a medium-high physics identity, with a mean of 3.8 (on a scale of 0 to 6). 
This mean is slightly higher than the middle point of the scale (3), showing that engineering students do not report a strong physics identity even when these majors are closely related to physics and science applications \cite{Tonso2006,Hazari2010}. 
The physics identity mean was different depending on the type of engineering major that students were enrolled in, ranging from a mean of 3.54 to a mean of 4.43 (see Table II). 
This difference between physics identity means could be related to the level of relationship that different engineering majors have with physics topics and applications, with majors such as mechatronic (4.43), chemical (4.27), and mechanical engineering (4.07) being the ones with students reporting the highest physics identity means; and majors like electric (3.54) and electronic (3.64) being the ones with students showing the lowest means (see Table II). 
These results suggest that having a high physics identity could be related to students’ major selection \cite{Hazari2010}, making them more likely to select a major related to physics topics and applications such as mechatronic and mechanical engineering if they develop a strong physics identity.

\subsection*{Physics Identity and its Relationship with Physics Courses Performance}
Most of the engineering students from the second sample completed their physics-related course with good results, with 166 students (83\%) getting a passing grade (equal to or greater than 70), and only 34 (17\%) failing it. 
The general grade average was 83.17 for the 200 students, and the grade average ranged between 74.53 and 88.23 within the different engineering majors (see Table II).  
Business management engineering was the major with the highest grade average (88.23), while mechanical engineering was the major with the lowest (74.53). 
These results may be biased by the level of difficulty that the physics-related courses from different engineering majors require to complete. 
While business management engineering requires an introduction to physics courses that only provide a basic review of physics topics like unit conversion, dynamic, and static, the mechanical engineering physics-related courses use advanced dynamic applications like the velocity of a cam or gears in mechanical devices. 
The difference between the complexity of the topics taught in each physics-related course depending on the major could have affected engineering students’ performance and grades in such courses, generating a more accessible way to high grades and good performance from the business management engineering students due to the lack of complex physics topics and applications in their physics course design. 

This discrepancy in the grade average of the physics-related courses due to the difference in the complexity of the topics taught in each major is also highlighted by the regression results shown in Table III and Table V. 
The multivariate linear regression that included the physics-related course as an independent variable showed that the physics-related course was the most significant factor correlated to students' grades in such courses (see Table III). 
Likewise, the logistic binomial regression that included the physics-related course as an independent variable showed that the only significant variable correlated to passing or failing the course was the type of physics-related course taken by engineering students depending on their major (see Table V). 
Aiming to thoroughly address the second objective of this research and better understand what other variables could be correlated to participants’ grades in their physics-related courses, the physics-related course independent variable was removed from both, the multivariate linear regression and the logistic binomial regression. 
The additional multivariate regression model showed that gender was a highly significant factor correlated with physics-related course grades (see Table IV), which is in line with the current theory that male students are more likely to get enrolled in physics-related courses and perform better in such courses \cite{Doucette2020,Herrera2020}. 
Another result of the multivariate regression models was that interest in physics-related activities was shown to be significantly correlated to physics-related course grades (see Table III and Table IV), suggesting that having an intrinsic interest in physic-related experiments and topics could have a positive effect on students grades in such courses \cite{Hsieh2012}. 
Additionally, the logistic binomial regression results showed that gender was the only significant factor correlated with passing their physics-related course (see Table VI). 
These results are once again supporting the gender gap shown in physics-related courses and careers favouring men in numbers and performance \cite{Kalender2019a}.  

\subsection*{Implications for Practice}
If STEM researchers and educators have access to appropriate tools to help them understand how their students feel when they are learning physics topics, then they could develop better strategies to help their students improve their physics abilities and performance in physics-related activities. Having a better understanding of the effects that experiencing different levels of physics identity could have on students' behaviour and performance in physics-related courses and majors would positively affect the way that STEM educators advise their students \cite{Jones2010}. 
This way, STEM educators around the globe would be able to design strategies to attract more young students to physics-related majors and to develop learning environments where students could practice and improve their physics abilities aiming to decrease the attrition rates in these careers. 
Additionally, having more instruments that address cultural and language barriers to collect data from Spanish-speaking Latino populations could help to solve the current underrepresentation or segregation of this population in physics-related careers \cite{Castaño2020}.

Finding better ways to generate an early interest in young girls in physics-related courses and activities could help STEM educators promote their physics identity development aiming to address the gender differences in these careers’ selection and persistence \cite{Hazari2007}. 
This is important because this gender gap is not based on intellectual differences between male and female students in the field of physics, as prior studies have found that physics female students usually perform as well as their male peers \cite{Hazari2007,Hyde2006,Sadler2001}. 
In the end, exposing minority students such as women to physics-related activities during their pre-college education could have a positive effect on the physics identity development of students with diverse backgrounds, helping to promote diversity and inclusion in the field of physics \cite{Singer2020}.

The physics identity mean (3.8) reported by engineering students in this study was considered as a medium-high value, suggesting that engineering students in general do not feel a strong physics identity even when some engineering majors are closely related to physics topics and applications. 
Once these results were analysed by different majors, the physics identity of engineering students from majors such as mechanics (4.07), chemical (4.27), and mechatronics (4.43) was higher than other engineering majors. 
This finding is important for STEM educators aiming to promote students’ interest in physics-related majors, since developing a high physics identity before college could help students to make the decision to pursue a STEM major that is closely related to physics topics and applications such as mechanical engineering, aerospace, robotics, and physics \cite{StilesClarke2018}. 
This could be more relevant for STEM educators trying to close the gender gap favouring the number of male students in physics-related majors \cite{Beasley2012}, making evident that female students could be more likely to show interest in such majors if they are presented with better opportunities to develop their physics identity in the early phases of their pre-college education.

Although physics identity was not a significant predictor of students passing or failing their physics-related course, the physics interest subconstruct was shown to be correlated with their physics-related course grades according to the linear regression models (see Tables III and IV). 
These results highlight the importance of promoting physics-related activities such as science fairs, physics demonstrations, and models building in middle school and upper high school aiming to develop an early interest in physics topics \cite{Adams2006,Lock2019}. 
Having a stronger physics identity coming from an intrinsic physics interest could help students to perform better in physics-related courses, and in consequence become more likely to consider a path in STEM careers. 
In the end, promoting physics-related activities aiming to increase students’ physics identity at an early age could help to address the gender gap in physics-related professionals if these activities are designed aiming to develop a stronger physics identity within female students in their pre-college preparation.     

\subsection*{Limitations and future work}
This research provides a good starting point for conducting physics identity research in Latino contexts and Spanish-speaking populations. 
However, future research using this adapted instrument for assessing physics identity needs to verify that its population culture and context is like the one described in this research, even when they talk the same language. 
Future data collections aiming to better understand the students’ physics identity effects on their grades and performance in physics-related courses should be focused on just one major with students taking the same course. This research analysed the physics identity of different engineering majors, and analysing students’ performance in physics-related courses with different difficulty levels could have created discrepancies in the statistic test results that may influence these research findings. 
Another future research direction should consider the analysis of the physics identity levels of physics majors’ students, and what was the effect of this physics identity in their decision to pursue such major. 
The physics identity data collected from physics majors’ students could be contrasted with data collected from students in different STEM majors to analyse possible coincidence and differences between this populations. 
This analysis could help STEM educators to design better strategies to attract more students to all STEM majors.    

\section{Conclusion}
This instrument adaptation could facilitate data collection for future research aiming to expand current physics identity literature within Spanish-speaking populations. 
This may be helpful to create more interest in physics identity studies from Latino cultures and contexts, as well as foster a more welcoming environment for the science research agenda for Latino researchers worldwide. 
This research stressed the importance of considering cultural and language adaptation when collecting data using translated and adapted instruments. 
It also emphasized the importance of validating and ensuring the reliability of datasets to improve the quality and trustworthiness of any research.

Developing and validating instruments to accurately assess students' physics identity is essential for STEM educators around the world to design and implement strategies that promote students' interest in physics. 
These strategies should focus on engaging students in physics-related activities such as science fairs and model building that explain physics concepts, aiming to develop a strong physics identity from an early age. Developing an intrinsic interest in physics-related activities during the pre-college education could motivate students to get involved in physics-related courses and enhance their performance in these courses. 
This, in turn, could generate greater interest among underrepresented groups, such as women, in physics-related careers.

\vspace{1cm}
\noindent \textbf{Acknowledgments.}
The authors state the acknowledgments information in relation to this study on the title page.

\noindent \textbf{Ethics Approval.}
Although this study pertains to low-risk, non-clinical research, no special permission from an ethics committee was deemed necessary. Nonetheless, the study strictly adhered to the WMA Declaration of Helsinki concerning ethical principles for medical research involving human subjects. Furthermore, compliance with the guidelines set forth by the German Research Foundation (DFG) and the German Psychological Society (DGP) was ensured.

\noindent \textbf{Informed Consent.}
All participants provided their informed consent by signing a document, ensuring the confidentiality of their information and that their data will be used exclusively for academic and scientific purposes. As such, all utilized information has been anonymized.

\noindent \textbf{Competing Interests.}	
The authors declare no competing interests.

\noindent \textbf{Author Contributions.}	
The authors contributed to this work as follows: Conceptualization by G. M.-S.; Data Curation by G. M.-S.; Formal Analysis by G. M.-S., O.I.G.P. and B.M. R.-L.; Funding Acquisition by O.I.G.P.; Investigation by G. M.-S., O.I.G.P., and B.M. R.-L.; Methodology by G. M.-S., O.I.G.P., and B.M. R.-L.; Project Administration by O.I.G.P.; Resources by G. M.-S., O.I.G.P., and B.M. R.-L.; Software by G. M.-S.; Supervision by O.I.G.P.; Validation by G. M.-S., O.I.G.P., and B.M. R.-L.; Visualization by G. M.-S. and O.I.G.P.; Writing – Original Draft by G. M.-S.; Writing – Review \& Editing by G. M.-S., O.I.G.P., and B.M. R.-L.

\noindent \textbf{Data Availability.}
The datasets generated or analyzed during the current study are available from the corresponding author on reasonable request. Due to privacy and ethical considerations, certain restrictions apply to the availability of these data.



%

\end{document}